\documentclass[journal=jctcce,manuscript=article]{achemso}

\usepackage{graphicx} % Required for inserting images
\usepackage{wrapfig}
\usepackage{amsmath}
\usepackage{comment}
\usepackage{booktabs}
\usepackage{rotating}

\usepackage{xr}

\makeatletter
\newcommand*{\addFileDependency}[1]{% argument=file name and extension
  \typeout{(#1)}
  \@addtofilelist{#1}
  \IfFileExists{#1}{}{\typeout{No file #1.}}
}
\makeatother

\newcommand*{\myexternaldocument}[1]{%
    \externaldocument{#1}%
    \addFileDependency{#1.tex}%
    \addFileDependency{#1.aux}%
}

\myexternaldocument{SI}

\usepackage{hyperref}
\hypersetup{
    colorlinks=true,
    linkcolor=blue,
    filecolor=magenta,      
    urlcolor=cyan,
    citecolor=blue,
    pdfpagemode=FullScreen,
    }
    
\author{Lindsey M. Whitmore}
\affiliation{Department of Chemical and Biological Engineering, University of Colorado Boulder, Boulder, CO 80309, United States}

\author{Yalda Ramezani}
\affiliation{Department of Chemical and Biomolecular Engineering, Ohio University, Athens, OH 45701, United States}

\author{Sumit Sharma}
\affiliation{Department of Chemical and Biomolecular Engineering, Ohio University, Athens, OH 45701, United States}

\author{Michael R. Shirts}
\affiliation{Department of Chemical and Biological Engineering, University of Colorado Boulder, Boulder, CO 80309, United States}
\email{michael.shirts@colorado.edu}

\title{Force switching and potential shifting lead to significant cutoff dependence in alchemical free energies}

\begin{document}

\maketitle
\begin{abstract}
The accurate treatment of long-range energy terms such as van der Waals interactions is crucial for reliable free energy calculations in molecular simulations. Methods like force switching, potential switching, potential shifting, and Ewald summation of van der Waals are commonly employed to smooth the truncation or otherwise manage these interactions at and beyond a cutoff distance, but their effects on free energy calculations are not always clear. In this study, we systematically explore the effects of these modifiers on the accuracy of free energy calculations using model systems: Lennard-Jones spheres, all-atom anthracene in water with GROMACS, and alkane chains in water with LAMMPS. Our results reveal that free energies of solvation using potential switching and particle-mesh Ewald summation of long-range Lennard-Jones are essentially independent of cutoff in solution, while force switching and potential shifting introduce cutoff-dependent behavior significant enough to affect the utility of the calculations.

\end{abstract}
\section{Introduction} 
Molecular simulations are powerful computational tools that can provide detailed insight into the behavior of molecular systems at the atomic level. Such simulations are widely used by researchers and have applications in drug discovery, material design, and the study of complex biophysical phenomena. By enabling the study of dynamic processes and subtle structural interactions that are difficult to capture experimentally, molecular simulations have become indispensable for researchers looking to understand and predict molecular behavior with high accuracy. A particularly important use of these simulations is to perform free energy calculations, which can be used to predict properties such as binding affinities and phase transitions. 

A key aspect in computationally efficient molecular simulations is accurate approximations in the treatment of long-range interactions, such as van der Waals (vdW) forces and electrostatic interactions,  which are fundamental to determining the physical properties and behaviors of molecules. van der Waals forces, which arise from induced dipole interactions between atoms, are especially crucial to include correctly in dense systems, like liquids and solution, as they can be responsible for a significant amount of cohesive energy in these dense phases. 

In most classical molecular simulations, van der Waals forces are modeled using pairwise potentials, with the Lennard-Jones (LJ) potential being by far the most commonly used functional form for describing these interactions, shown in Eq.~\ref{eq:lj}:
\begin{equation}
  U_{ij}(r) = 4\epsilon_{ij} \left[\left(\frac{\sigma_{ij}}{r}\right)^{12} - \left(\frac{\sigma_{ij}}{r}\right)^6\right]
  \label{eq:lj}
\end{equation}

However, the computational cost of calculating these interactions for all particle pairs in a system increases approximately quadratically  as the system size grows. To manage this computational expense, a cutoff distance is typically applied, beyond which interactions are either neglected or approximated. Cutoff distances for vdW interactions usually range from 0.8 to 1.4 nm, depending on the system, the precision required, and the defaults for the force field and simulation engine used.  The effect of the dispersion energy beyond this cutoff can be estimated analytically very accurately for systems that become homogeneous beyond the cutoff with a nearly costless correction~\cite{Shirts:2007:J.Phys.Chem.B}. In systems that are more inhomogeneous beyond the cutoff, more complex solutions such as periodic lattice summation methods including Ewald and particle mesh approaches may be required to obtain accurate thermodynamic properties that are independent of the cutoff~\cite{Wennberg:2013:J.Chem.TheoryComput.,Wennberg:2015:J.Chem.TheoryComput.,intVeld:2007:TheJournalofChemicalPhysics}.  

Abruptly truncating these interactions at the cutoff distance can lead to artificial forces and energy discontinuities, disrupting the consistency of the numerical integration of the simulation. When particles cross the cutoff distance, the sudden removal of forces can cause unphysical artifacts, destabilization of structures, heating of the system, or errors in calculated properties such as diffusion coefficients and phase behaviors.~\cite{Shirts:2007:J.Phys.Chem.B}  Both shifting of the potential function and incorporating tail corrections have been shown to significantly impact the phase diagram of Lennard-Jones fluid.~\cite{smit1992phase} Errors in force calculation attributed to spherical cut-off are exacerbated near interfaces due to inhomogeneous distribution of particles.~\cite{wang2012error} Moreover, the tail of the potential that is often truncated in simulations strongly affects the calculation of surface tension.~\cite{nijmeijer1988molecular}.  Although the abrupt truncation of Lennard-Jones potentials causes substantially less error than the electrostatic interactions because of the significantly smaller in magnitude $1/r^6$ potential versus the $1/r$ potential, and can be mitigated to substantial extent with a thermostat, simulations with abrupt van der Waals truncation will not conserve energy in a constant energy simulation, for example.  

In an effort to mitigate the inaccuracies introduced by abrupt cutoffs, various techniques have been developed to smooth this transition in the pairwise potential energy. These approaches aim to ensure that the interactions at the cutoff distance are treated more smoothly, minimizing the disruptive effects of truncation while maintaining computational efficiency.
In particular, GROMACS supports three alternatives to handing this discontinuity when using \texttt{vdwtype = cutoff}: potential switching, potential shifting, and force switching. 

Although particle mesh Ewald (PME)~\cite{PME_1993} has historically been used for handling long-range electrostatic interactions, it can also be applied to van der Waals forces as an alternative to cutoff-based truncation~\cite{PME_1995,Wennberg:2013:J.Chem.TheoryComput.}. Instead of truncating interactions at a fixed cutoff, PME employs Ewald summation, dividing interactions into short-range components, which are computed directly in real space, and long-range components, which are evaluated in reciprocal space, in the case of PME interpolated along a grid so that the fast Fourier transform can be used~\cite{PME_1995}. This method ensures continuous and differentiable treatment of interactions while reducing artifacts associated with abrupt truncation. While PME introduces additional computational complexity due to the need for grid-based interpretation and Fourier transformation, it eliminates the need for explicit switching functions. GROMACS supports PME calculations for the $1/r^6$ dispersion interaction.  Switching functions can be used for the real space portion, but because the real space van der Waals energy is by design only a small fraction (0.1\% or less) of the total van der Waals energy at the cutoff, the effect is negligible.  

Mathematically, these different types of cutoff can be described as follows:

\textit{Potential switching} gradually scales the potential energy to zero between the full potential and zero, starting at some user-specified shorter switching distance and ending at the cutoff distance. The potential-switch function as implemented in GROMACS \cite{abraham_2024_13457083} is

\begin{equation}
    S_v(r) = 1 - 10\left(\frac{r - r_s}{r_c - r_s}\right)^3 + 15\left(\frac{r - r_s}{r_c - r_s}\right)^4 - 6\left(\frac{r - r_s}{r_c - r_s}\right)^5
\end{equation}
Where $r_c$ is the cutoff and $r_s$ is the start of the switching distance. This satisfies the conditions that $S_v(r)=1$ at $r_s$ and $S_v(r) = 0$ at $r_c$, as well as both the first and second derivative of $S_v$ being zero at both $r_s$ and $r_c$.

Then the truncated potential $U_{trunc}(r)$ between two particles with Lennard-Jones interactions is: 
\begin{equation}
    U_{trunc}(r) = 
    \begin{cases}
    U_{lj}(r) &  r < r_s \\
    S_v(r)U_{lj}(r) & r_s \leq r < r_c \\
    0 & r_c \leq r 
    \end{cases}
\end{equation}

The chain rule can be used to calculate the force over the range, which will be both continuous and first differentiable, since the potential is twice differentiable.  The switching force results in a slight ``bump'' in the pairwise force between $r_s$ and $r_c$, making it somewhat larger in magnitude than the force approaching $r_c$.  Although this effect is if the switch is of moderate (0.05 to 0.1 nm) width and the cutoff it at moderate (0.9 to 1.0 nm) distance, is not entirely physical.

\textit{Potential shifting}, in contrast, applies a constant offset to each pairwise potential energy term equal to the value of the energy at the cutoff $V(r_c)$. Shifting the potential upwards ensures continuity in the potential energy function at the cutoff distance, and ensures that the forces are identical to the original potential up to $r_c$. The second derivative is still discontinuous at that point, creating small errors in molecular dynamics integration, but the effect is generally negligible in almost all realistic situations. 

\textit{Force switching} involves algebraically modifying the force between pairs of particles to satisfy desired conditions of the force at or beyond the boundaries; of course, modifying the force affects the potential as well. A switching function gradually reduces the forces to zero between the switch and the cutoff distance, ensuring that the forces remain continuous and monotonic at all points. This method aims to eliminate the non-monotonic force changes that appear in potential switching. 

For potentials of the form
\begin{equation}
    U_{\alpha}(r) = r^{-\alpha} \nonumber 
\end{equation}

Then the force is:
\begin{equation}
    F_{\alpha}(r) = \alpha  \cdot  r^{-(\alpha + 1)} \nonumber
\end{equation}
The switched force can be represented as:
\begin{equation}
    F_{s(r)}
    \begin{cases}
    F_{\alpha}(r) &  r < r_s \\
    F_{\alpha}(r) + S_F(r) & r_s \leq r < r_c \\
    0 & r_c \leq r 
    \end{cases}
\end{equation}
The switching function in GROMACS\cite{abraham_2024_13457083} is:
\begin{equation}
    S_F(r) = A(r-r_s)^2 + B(r-r_s)^3
\end{equation}
and satisfies the requirements that $S_F(r_s)=S_F^\prime(r_s)=0$, and at $r_c$, $S_F$ and $S_F^\prime$ are equal to the force and it's derivative, thus canceling them  out.  The force can be integrated to give the shifted potential in GROMACS \cite{abraham_2024_13457083}, implemented as:
\begin{equation}
    U_s(r) =  \frac{1}{r^{\alpha}} - \frac{A}{3}(r-r_s)^3 - \frac{B}{4}(r-r_s)^4 - C
\end{equation}
Where
\begin{equation}
A = \frac{-\alpha (\alpha + 4)r_c - (\alpha + 1)r_s}{r_c^{\alpha+2} (r_c - r_s)^2} \nonumber
\end{equation}
\begin{equation}
    B = \frac{\alpha (\alpha + 3)r_c - (\alpha +1)r_s}{r_c^{\alpha+2} (r_c - r_s)^3} \nonumber
\end{equation}
\begin{equation}
    C = \frac{1}{r_{c}^{\alpha}} - \frac{A}{3}(r_c - r_s)^3 - \frac{B}{4}(r_c-r_s)^4 \nonumber
\end{equation}

For the Lennard-Jones potential, we add together the potential function with $\alpha=12$ term, subtracted by that with the $\alpha=6$, all multiplied by $4\epsilon$.

%\item \textit{Force shifting} modifies the forces directly,  without altering the shape of the potential function. A constant is added to the forces so they reach zero at the cutoff, ensuring continuity and preventing destabilization.
%\begin{equation}
%    F_s (r)= 
%    \begin{cases}
%        F(r) & r < r_s \\
%        \frac{\alpha}{r^{\alpha +1}} + %A(r-r_s)^2 + B(r-r_s)^3 & r \leq r_c \\
%        0 & r \geq r_c \\
%    \end{cases}
%\end{equation}
%end{itemize}

As mentioned before, the omitted energies due to modifications of the van der Waals terms can either be neglected or approximated.  For switched pairwise potentials, the most common correction is to assume that the radial distribution function $g(r)=1$ for all types of particle pairs outside the switched distance, and analytically integrate the van der Waals potential energy times 1 minus the switching function out to infinity for all pairs, and can be done once for an entire simulation in the NVT ensemble. Because it deals with the tail of the vdW potential,  it is sometimes called a "tail correction".  This correction can also be applied to the pressure by integrating out the contribution to the virial~\cite{Shirts:2007:J.Phys.Chem.B}, though it needs to be applied every step. This analytical approximation is very accurate for systems, such as simple fluids, that become homogeneous outside of the cutoff distance, and this approximation is nearly costless.~\cite{Shirts:2007:J.Phys.Chem.B} This correction is only needed for $r>r_{switch}$, as the potential is unmodified over the range $r<r_{switch}$.   For systems such as proteins in water, or a ligand bound to a protein, then there is some anisotropy, and the correction is not quite as accurate, especially at lower common cutoffs like 0.9 nm, and slightly more care may need to be applied to get quantities like protein-ligand binding energies correct~\cite{Shirts:2007:J.Phys.Chem.B}. 

For potential modifiers with a shift that is subtracted off of the potential, including both potential shifted and force shifted potentials, such a correction is more complicated, because the potential is modified at all $r$, and $g(r)$ is very different from one inside the cutoff.  However, the correction which GROMACS applies in the case of potential shifting is ${\frac{\rho}{2}}\int_0^{r_v} V_s g(r) 4 \pi r^2  dr = \frac{\rho V_s}{2} \int 4 \pi r^2 g(r) dr = \frac{V_s}{2} N_c$, where $V_S$ is the value of the shift at the cutoff, a constant, and $N_c$ is the average number of particles within the cutoff. Assuming that $g(r)\approx 1$ at the cutoff and the system is homogeneous, this will still be a reasonable approximation, as $\frac{N_C}{\frac{4}{3}\pi r_C^3}$ will be close to the average density of the system. 

In systems such as membrane bilayers that are more inhomogeneous beyond the cutoff point, more complex solutions such as periodic lattice summation methods including Ewald and particle mesh approaches may be required to obtain accurate thermodynamic properties independent of the cutoff~\cite{Wennberg:2013:J.Chem.TheoryComput.,Wennberg:2015:J.Chem.TheoryComput.,intVeld:2007:TheJournalofChemicalPhysics}.  

In this paper, we are primarily interested in the effect of these truncation and approximation methods on calculated free energies. The most common way to treat the removal of Lennard-Jones interaction sites in alchemical free energy calculations is by using a soft core function, which reduces the height of the repulsive core down from infinity as a function of $\lambda$. In the soft core functional form of Beulter et al.~\cite{Beutler:1994:ChemicalPhysicsLetters}, as implemented in GROMACS, the van der Waals energy is modified for free energy calculations connecting states $A$ and $B$ as:
\begin{equation}
    V_{sc}(r) = (1 - \lambda)^n V^A(r_A)+ \lambda^n V^B(r_B)
\label{eq:soft_core}
\end{equation}
%where $V^A$ and $V^B$ are the hard core potentials, and:
\begin{equation}
    r_A = (\alpha {\sigma}_B^6{\lambda}^p)^\frac{1}{6} , \nonumber
\end{equation}
\begin{equation}
    r_B = (\alpha{\sigma}_B^6(1-\lambda)^p + r^6)^\frac{1}{6} ,
    \nonumber
\label{eq:soft_core_LJ}
\end{equation}
$\alpha$ is the Lennard-Jones soft-core parameter, $n$ is a positive exponent, $p$ is the soft-core $\lambda$ power, and $V^A$ and $V^B$ are the full vdW potentials in states A ($\lambda$ = 0) and B ($\lambda$ = 1). In the case of simulations carried out in this study, our B state has no interactions between solute and solvent, so $V_B = 0$, and the transformation is the decoupling of the solute molecule from solvent, giving us a free energy of solvation. 

There are other proposed soft core functional forms~\cite{Gapsys:JCTC:2012,Naden:JCTC:2014,pham_identifying_2011,Pham:JCP:2012,Naden:JCTC:2015,Li:JCTC:2020}, which have somewhat improved numerical properties, but they have similar behavior as a function of $\lambda$.

We can see from this functional form that the energy at the cutoff in the unmodified potential $V(r_c)$, which is subtracted off in both force switching and force shifting, is $\lambda$-dependent. Thus, the potential shifting or force-switching term is $\lambda$-dependent in a way that is more complicated to incorporate into the calculation of potential energy differences at different $\lambda$, used in MBAR or BAR, or the derivative of the potential with respect to $\lambda$ as used for TI.  In contrast, for potential switching, the $\lambda$ dependence on the switch is directly incorporated because it is solely a multiplicative factor. 

We have verified for GROMACS that this cutoff dependence is properly calculated in the calculation of free energy differences for all cutoff types for the inputs of BAR and MBAR calculations. This was done by comparing the energy differences between two different $\lambda$ states that share the same configurations when generated in two different ways. The first is taking the potential energy differences produced directly by the free energy code within a single simulation. The second is carrying out simulations at one value of $\lambda$, and then using the GROMACS \texttt{mdrun -rerun} functionality to calculate the potential energy differences when using different values $\lambda$ with each set of trajectories.  Both methods agreed for all calculation types described in the paper, indicating the correct implementation of free energy differences.  This comparison demonstrates there is not an issue of calculating the $\lambda$ dependence of the potential energy cutoff when computing the quantities needed for free energy calculations during the simulation run.  A script to carry out this comparison is provided at \href{https://github.com/shirtsgroup/Shift_switch_free_energy}{https://github.com/shirtsgroup/Shift\_switch\_free\_energy}. Calculation of $\langle \frac{dU}{d\lambda}\rangle$ as used in TI was not directly verified, but TI calculations agreed with MBAR for all simulations, indicating the TI implementation was also likely correct to within statistical precision.

Despite their widespread use, the exact impact of these different choices to modify the dispersion interactions has not been thoroughly quantified. 
In this study,  we systematically explore the differences in the free energies calculated with these approaches using simple model systems.  We first examine the free energy as a function of distance between two Lennard-Jones spheres as a function of coupling variable $\lambda$ using soft core interactions to observe the differences in the potential as a function of $\lambda$. We next assess the effect of these cutoff modifications on the free energy calculations using a model of anthracene in water over a range of different cutoff distances.  
Finally, we also note that this observation about the difference in free energies between different cutoff types is not exclusive to GROMACS.  LAMMPS also supports potential-shifting of the van der Waals functions, and we show results of differences in hydration free energy of alkanes depending on which truncation method is used.

We note that when performing this study, we initially identified an additional error in the implementation of force-switching combined with free energy calculations in GROMACS, apparently existing since the introduction of the force-switching modifier. Upon notification, this error was rapidly fixed by the developers and is correct starting from the recently released version 2024.3, the version of GROMACS used for all calculations in this study.  We do not determine the effect of any previous errors in the force-switch modifier on the free energies, only on the current correct implementation of switching and shifting functions.

\section{Methods}
\subsection{Lennard-Jones Spheres}
\subsubsection{System Set-up}
The system consisted of two Lennard-Jones spheres, with interaction parameters of $\sigma$ = 1.0 nm and $\epsilon$ = 1.0. The particles were separated along the x-axis, with their y- and z- coordinates fixed at 0. Initial particle distances were 0.2 nm to ensure enough separation to prevent out-of-scale energies. Particle positions were modified iteratively by increasing the x-separation by 0.001 nm.

For each particle configuration, point calculations were performed across 11 $\lambda$ states, spaced equally from $\lambda$ = 0 to $\lambda$ = 1, using the soft-core interaction of Beutler et al.~\cite{Beutler:1994:ChemicalPhysicsLetters} 
In equation ~\ref{eq:soft_core_LJ}, $p = n = 1$, and $\alpha = 0.5$, the GROMACS defaults, were used. 
The potential energy was evaluated with each GROMACS vdw-modifier with switching distance of 1.75 nm and cutoff distance of 2.0 nm.
A Python script (provided at \href{https://github.com/shirtsgroup/Shift_switch_free_energy}{https://github.com/shirtsgroup/Shift\_switch\_free\_energy}) was used to automate the process of modifying particle positions and preparing input files for each iteration. GROMACS XVG format files were subsequently converted to CSV format for ease of data processing and visualization.

\subsection{Running Free Energy Simulations}
\subsubsection{GROMACS simulations}
All molecular dynamics (MD) simulations were performed using GROMACS 2024.3. Simulations were run for each GROMACS vdw-modifier, implemented with the \texttt{vdwtype=Cut-off} and \texttt{vdw-modifier=
Force-Switch}, \texttt{Potential-Shift}, or \texttt{Potential-Switch} GROMACS mdp options, at multiple cutoff distances. The simulation setup and parameters were kept consistent across all systems, except for the vdw modifiers and cutoff values.

A test system consisting of a simplified model of anthracene, solvated in TIP3P water, was used for all GROMACS free energy calculations. Anthracene was chosen to be relatively large, to make sure any cutoff effects are clear and obvious. Also, because it has no internal rotatable bonds resulting in slow internal degrees of freedom, simulations converged rapidly.  To further simplify the problem we used a united atom model of anthracene, with single particles representing the C-H moieties, and with all zero partial charges.  The initial configuration was obtained from a previously published study~\cite{Hsu:2024:J.Chem.TheoryComput.}. Energy minimization was performed using the steepest descent algorithm, followed by equilibration performed in the NVT ensemble for 100 ps at 300 K using the v-rescale thermostat and tau\_t of 0.1 ps.  

All production simulations were carried out in the NVT ensemble using the velocity-Verlet integrator for 200 ns of simulation using a time step of 2 fs. For each modifier, the cutoff distances were varied systematically across the range of 0.75 nm to 1.2 nm. Specifically, the cutoff conditions used (listed as switching distance $r_s$ and cutoff $r_c$) were 0.75--0.8 nm, 0.85--0.9 nm, 0.95--1.0nm, 0.8--0.9 nm, 0.9--1.0 nm, and 1.1--1.2 nm. 15 $\lambda$ intermediates (0,00, 0.09,0.180, 0.300, 0.420, 0.495, 0.570, 0.625, 0.680, 0.720, 0.810, 0.860, 0.930, 1.00), roughly chosen to minimize total variance were used to turn off the van der Waals interactions, using the Beulter soft-core functional form.~\cite{Beutler:1994:ChemicalPhysicsLetters}. 

The simulations used the expanded ensemble framework, using Metropolized Gibbs Monte Carlo moves~\cite{choderaReplicaExchangeExpanded2011} to change the $\lambda$ state during the simulation, with moves attempted every 100 MD steps. Rather than being determined self-consistently, we used fixed weights for each state (values in Table~\ref{SI:tab:ee_weights}) which resulted in approximately even sampling for all states for this problem. In all simulations, states sample multiple times between $\lambda=0$ and $\lambda=1$ during each simulation. Since the different cutoffs resulted in somewhat different free energies and therefore different weights for equal sampling, simulations did not visit all $\lambda$ states exactly equally, but this is not required for free energy calculations, as long as all states have some visitation.  Because the weights were not updated during the simulations, each state rigorously samples the proper Boltzmann distribution, and BAR and MBAR can be applied with no approximation.

The simulations performed to evaluate the effect of PME were prepared using the same general protocol and a cutoff range of 0.75--0.9 nm with ewald-rtol set to \( 1 \times 10^{-6} \) and ewald-rtol-lj set to 0.001 to control the relative error of the real-space term at the cutoff for electrostatic and dispersion interactions, respectively.
Free energies were calculated using MBAR, BAR, and TI using the alchemlyb~\cite{Wu2024-alchemlyb} and  pymbar packages \cite{michael_shirts_2024_10849928-pymbar}.  Ten replicates were performed for each set of cutoffs and vdW-modifiers, and uncertainties are calculated as the standard error in the mean over these 10 replicates. All input files are available on github at: \href{https://github.com/shirtsgroup/Shift\_switch\_free\_energy}{https://github.com/shirtsgroup/Shift\_switch\_free\_energy}.

\subsubsection{LAMMPS simulations}

Hydration free energy of n-alkanes [methane to eicosane, $C_{20}$] were computed with shifted and unshifted potentials using the TIP4P/2005 (a 4-site model) water model and TraPPE-UA force field for alkanes with the alkane and water-oxygen cross-interaction Lennard Jones well depth ($\epsilon$) modified to obtain a better match between the hydration free energies from simulations and experiments.~\cite{ashbaugh2011optimization} This modified force field is called HH-alkane. The simulations were performed using LAMMPS (version March 2021) using 64 cores on a 64-core AMD EPYC 7742 processor. Depending on the system size, the simulation speeds vary from 0.67 ns/h to 1.67 ns/h. Experimental hydration free energies are known up to n-octane, beyond which we employ the group contribution approach for comparing the simulation results with the experiments.~\cite{Cabani1981} As per the TraPPE-UA force field, the spherical cutoff for alkane-alkane and alkane-water interactions was set at 1.4 nm, for both the shifted and unshifted potentials. A spherical cutoff of 1.0 nm was used for both Lennard-Jones and Coulombic interactions between water molecules. Long-range electrostatics were calculated using particle-particle particle mesh Ewald summation (PPPM or P3M) with a tolerance of $10^{-4}$. The simulations were conducted in the isothermal-isobaric ensemble (constant number of particles \textit{N}, temperature \textit{T}, and pressure \textit{P}) with \textit{T} = 300 K and \textit{P} = 1 bar. The \textit{T} and \textit{P} were maintained using the Nos\'{e}-Hoover thermostat and barostat, respectively. The simulation time step was set to 1.5 fs. The size of the simulation system was chosen to ensure that the alkane molecule does not interact with its periodic image. The system sizes are listed in Table \ref{SI:tab:alkane_water}.

Hydration free energies were computed using the free energy perturbation (FEP) methodology. In this methodology, the interactions between water-oxygen and alkane united atoms are modeled via a soft potential given by equation \ref{eq:soft_core_LJ}.The value of $\alpha$ was set to 0.5 and $n = p = 2$. Differences in these parameters between software programs can change the pathway in interaction potentials traced as $\lambda$ varies and the statistical efficiency of the calculations~\cite{pham_identifying_2011,Pham:JCP:2012}, but in the limit of sufficient sampling the end point free energies will be equal.

$\lambda$ is varied from 0 to 1 in a stepwise manner. We considered 40 windows with changes in $\lambda$, $\Delta\lambda = 0.025$. In each window, simulations were performed for 6 ns, and the last 4.5 ns were used to perform the ensemble averages. Accuracy of the FEP simulations was determined by also calculating the free energy change associated with the reverse path where $\lambda$ is varied from 1 to 0 with the same $\Delta \lambda$. The reported hydration free energy values are the average of the forward and the reverse paths (see tabulated results in~\ref{tab:SIlammps_free_energies}). The difference in the hydration free energy estimate between the forward and the reverse paths is considered as error. 

%\begin{equation}
%    V(\lambda, r) = \lambda^n4\epsilon \left[\left(\alpha (1 - \lambda)^2+\left(\frac{r}{\sigma}\right)^6\right)^{-2} - \left(\alpha(1-\lambda)^2+\left(\frac{r}{\sigma}\right)^6\right)^{-1}\right]
%\label{eq:soft_core_LJ}
%\end{equation}

Hydration free energy is the change in the free energy when a solute enters the aqueous phase from the ideal gas state and is calculated using the equation \ref{eq:fep}. We calculated the hydration free energy with unshifted potentials with tail corrections for energy and pressure and with shifted potentials as described above. 
% already defined above, so reducing verbiage.
%Here, chifted potential refers to the potential that is truncated at the spherical cutoff and the entire potential function is shifted by the value of the potential at the cutoff distance, so that the potential function becomes zero at the cutoff.  

\begin{equation}
\Delta G = -k_BT \sum_{i=0}^{n-1} \ln\left(\frac{\langle V(x) \exp\left(-\frac{U(\lambda_{i+1},x)-U(\lambda_{i},x)}{k_BT}\right)\rangle _{\lambda_i}}{\langle V(x) \rangle _{\lambda_i}}\right)
\label{eq:fep}
\end{equation}

Input files and scripts for analysis for the LAMMPS free energy workflow are provided at:
\href{https://github.com/shirtsgroup/Shift_switch_free_energy}{https://github.com/shirtsgroup/Shift\_switch\_free\_energy}.

\section{Results}

\subsection{The interaction between cutoffs and alchemical coupling parameter}
\begin{figure}
    \centering
    \includegraphics[width=1.0\linewidth]{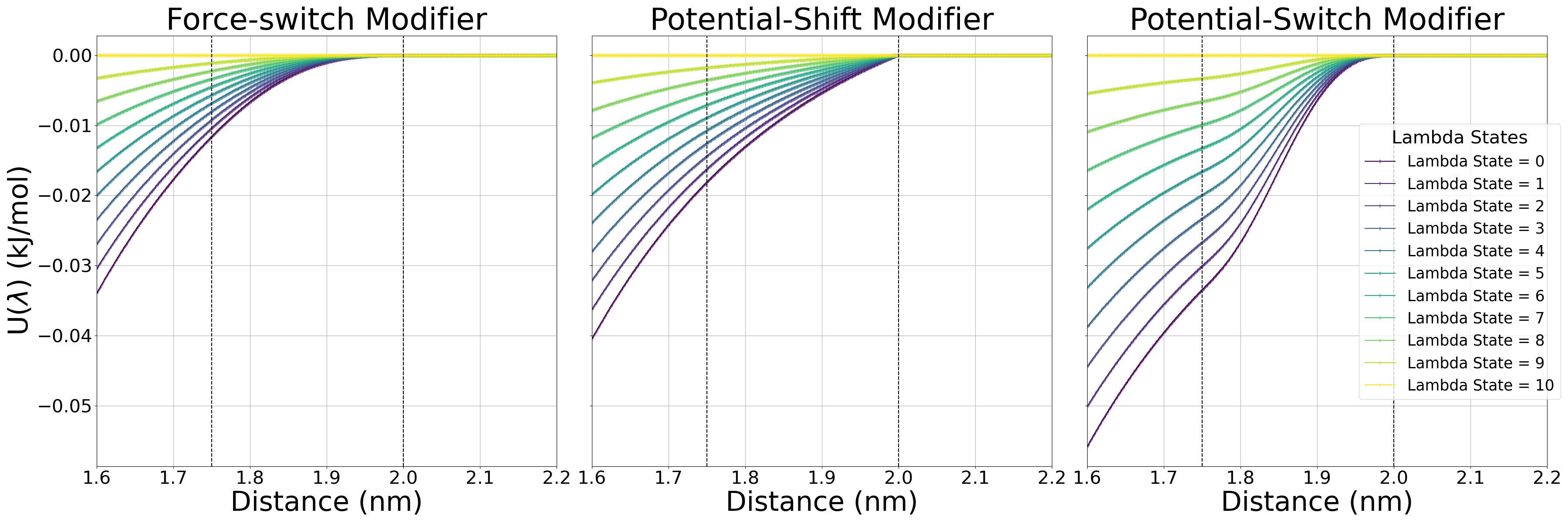}
    \caption{Potential energy U($\lambda$) between two Lennard-Jones spheres as a function of inter-particle distance for the three van der Waals modifiers examined in this study. The potential energy (U($\lambda$)) is plotted  as a function of the distance (nm) between the Lennard Jones spheres at different values of $\lambda$,  The vertical lines mark the switching distance (1.75 nm) and the cutoff distance (2.0 nm). The three panels show the force-switch (left), potential-shift (center), and potential-switch (right) GROMACS modifiers. 
    \label{fig:twospheres}}
\end{figure}

To understand the behavior of each modifier and its effects on U($\lambda$), we examined potential energy profiles between two LJ spheres as a function of distance for a series of equally spaced 11 $\lambda$ states using the three vdW modifiers (Fig. ~\ref{fig:twospheres}). As expected, at all $\lambda$, for all modifiers, the potentials and forces go smoothly to zero.  However, we see that the shifts of the force-switch and potential-shift modifiers depend on $\lambda$ inside the switching region ($r < r_{rswitch}$), in contrast to the potential-shift potential.

\subsection{The effects of modified van der Waals potentials on solvation free energies}

\begin{figure}
    \centering
    \includegraphics[width=0.9\linewidth]{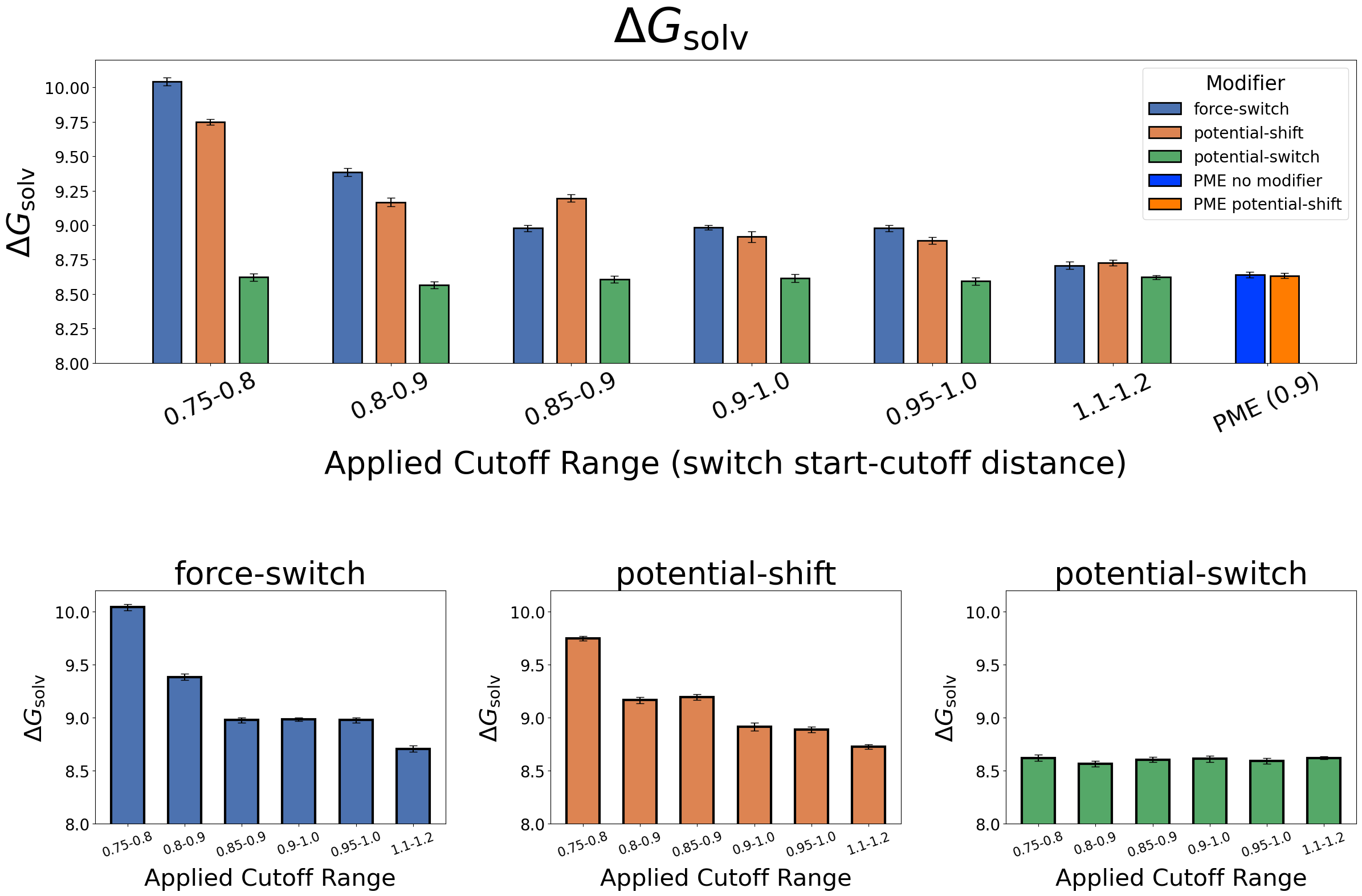}
    \caption{Free energy of solvation ($\Delta G_{solv}$) as a function of applied cutoff range in kJ/mol. The top panel shows $\Delta G_{solv}$ across cutoff ranges for the force-switch (dark blue), potential-shift (orange), and potential-switch (green) modifiers, as well as PME, evaluated using the Multistate Bennett Acceptance Ratio (MBAR) (comparison to other methods is shown in SI figure~\ref{SIfig:all_methods_FE}). $\Delta G_{solv}$ calculated with force-switch and potential-shift modifiers has significant cutoff dependence, whereas the $\Delta G_{solv}$ calculated with the potential-switch modifiers is essentially cutoff dependent. $\Delta G_{solv}$ calculated with different PME variations (light blue, orange) is consistent with the potential-switch modifier results . The bottom panels show the free energy contributions for each modifier individually for ease of comparison. Error bars represent the standard error of the mean for $\Delta G_{solv}$ values over 10 replicates for each modifier across the applied cutoff ranges.}
    \label{fig:freeenergies}
\end{figure}

The free energy of solvation, $\Delta G_{solv}$, for each long range vdw treatment using different cut-off ranges is plotted in Figure \ref{fig:freeenergies}. 

We find that the $\Delta G_{solv}$ of the simulations with the potential-switch modifier were statistically unchanged with cutoff. In contrast, the potential-shift and force-switch exhibit significant variability across the cutoff ranges, but converge towards the $\Delta G_{solv}$ obtained using the potential-switch modifier as the cutoff range increases and the magnitude of any modifier decreases.  The solvation free energy obtained using PME is similar to the values computed using potential-switch, further supporting that potential switching provides a reasonable approximation of long-range dispersion contributions at sufficiently long cutoffs.

\subsection{Speed of different cutoff methods} To analyze the speed of different methods to treat the long range van der Waals interactions, we evaluated the speed in ns/sec to carry out 10 replicates of the anthracene free energy calculations with potential shifting, potential switching, force switching, and van der Waals PME, in the last case with both cutoffs and potential shifting of the real space component. All simulations were run using GROMACS 2024.3 on nodes equipped with AMD Milan CPUs (64 cores, 3.8 GB RAM per core, 32 MB L3 cache) running RHEL 8.4. Each job was submitted via SLURM and executed on a single node using 12 CPU cores via OpenMP parallelization by specifying -nt 12 in the gmx mdrun command. GPU acceleration was not used. This level of parallelization was consistent across all replicates. Speeds in nanoseconds per day were collected directly from GROMACS log files.  
Uncertainties in speed on a run-to-run basis were relatively high, because of the practical variance in a high-speed computing platform. 
We found that potential shifting was the fastest, with potential switching slower by 5\% $\pm$ 4\% (163 $\pm$ 6 ns/day versus 155 $\pm$ 2 ns/day).  Force switching (148 $\pm$ 10 ns/day) was slower than potential shifting by 9\%$\pm$7\% , whereas PME without any modifier (98 $\pm$ 7 ns/day) was slower by 40\% $\pm$ 5\%, and PME with real space potential shift (119 $\pm$ 7 ns) by 27\% $\pm$ 6\%.  Given the almost negligible difference between potential shift and potential switch, this further solidifies the desirability of potential switching for free energy calculations. We note that such performance measures are highly dependent on system size, simulation parameters, and the computational resources and paralleization used, so these results may not be directly applicable, but agree broadly with the previous examinations of timing differences.
Full timing data found in the supporting in formation in Table~\ref{tab:SIpme_speed}.  

\subsection{Hydration free energies using LAMMPS}

\begin{figure}
    \centering
    \includegraphics[width=0.9\linewidth]{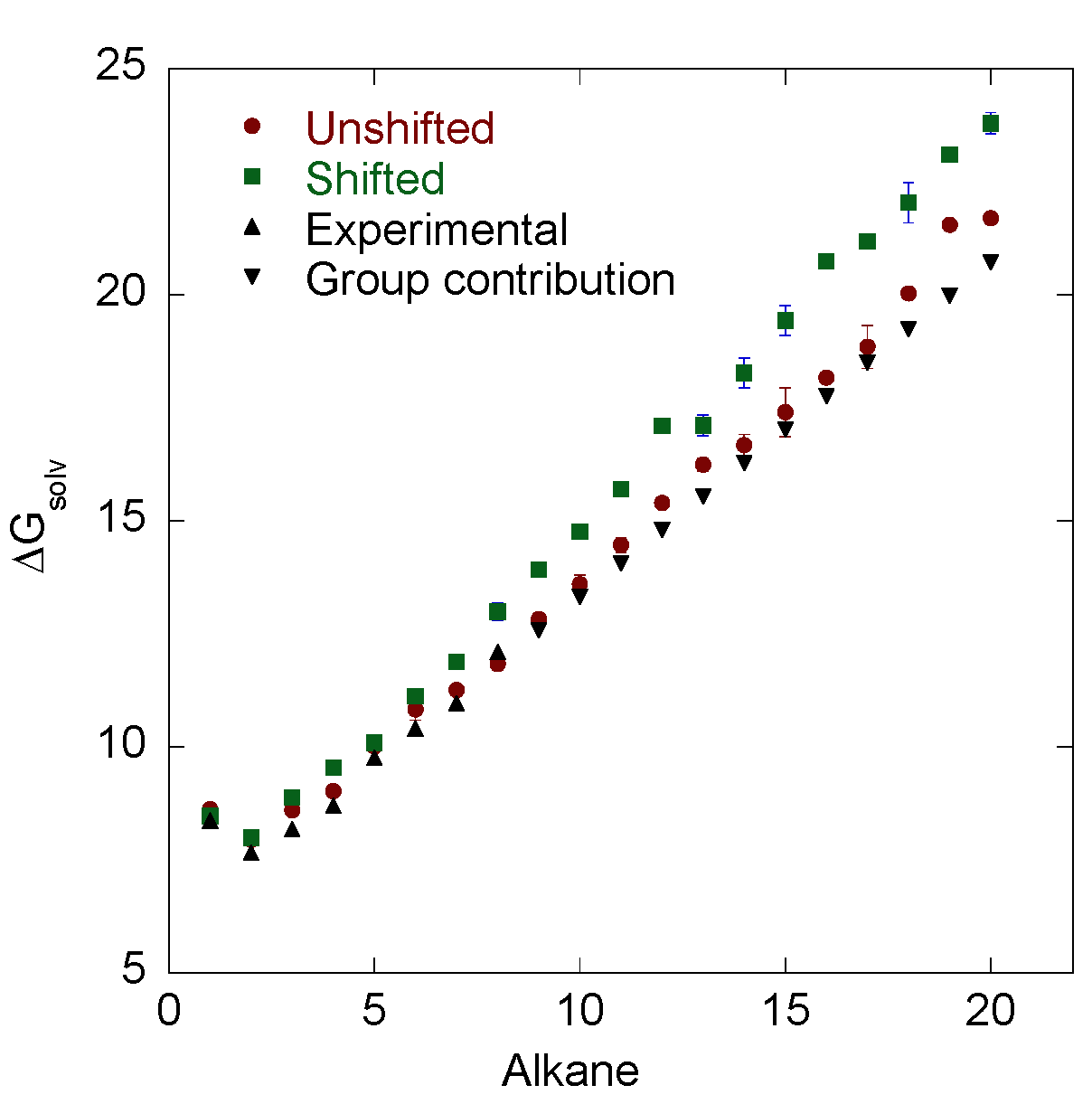}
\caption{\label{fig:lammps}Hydration free energies of the shifted (green) and the unshifted (red) potentials using the TIP4P/2005 water model and the HH-alkane model for alkanes in kJ/mol compared to the experimental values and those obtained using the group contribution method (black).  There is a significant, size-dependent, difference between 
shifted and unshifted results. %The results show that the hydration free energies from the unshifted potential are closer to the experimental / group contribution values. The maximum deviation is 1.57 kJ/mol for C$_{19}$. In contrast, the estimates from the shifted potential show a systematic deviation of, on average, 0.15 kJ/mol/carbon from the experimental / group contribution values.
}
\end{figure}

We compared the hydration free energy of n-alkanes ($C_1$ – $C_{20}$) calculated with shifted potentials with those from unshifted potentials with tail corrections for TIP4P/2005 (a 4-site model) water model and HH-alkane model for alkanes using LAMMPS. We also compare the calculated hydration free energy with experimental values and the group contribution method(Fig~\ref{fig:lammps}). 

We find that the hydration free energies from the shifted potential systematically deviate from the unshifted potential by approximately 0.1 kJ/mol/carbon. Except for methane, the hydration free energies are over-predicted by the shifted potential as compared to the unshifted potential. This is not surprising as the attractive energy is reduced by the shifted amount in the shifted potential. For the unshifted potential, the difference from the experimental / group contribution values is, on average, 0.31 kJ/mol up to decane ($C_{10}$). Beyond decane, the difference becomes larger. The maximum difference is observed for $C_{19}$, equal to 1.57 kJ/mol. On the other hand, hydration free energy from the shifted potential shows a systematically increasing deviation from the experimental / group contribution values at the average rate of 0.15 kJ/mol/carbon. Clearly, hydration free energies from unshifted potential are much closer to the experimental values.  However, the exact difference in comparison to experiment is less of a concern; the large difference between the shifted and unshifted treatments with a 1.4 nm cutoff means that it is challenging to ensure repeatability between different treatments between studies or in force field development. Table~\ref{tab:SIlammps_free_energies} provides hydration free energy values and differences for alkanes using both approaches.

\section{Discussion}

For smaller molecules and larger cutoffs, these distance-dependence differences may be more negligible. We also note that for drug binding affinities, one is looking at the difference in solvation free energies between the protein and ligand environment, both of which will have cutoff errors that are often of the same magnitude, so the difference in binding affinities will be lower than the differences for solvation free energies.

In many studies, cutoffs between 0.9 and 1.0 nm are common, which for anthracene would result in errors up to 0.6 kJ/mol. In studies at medium to low precision, the errors in free energy as a function of cutoff treatment observed in this study are therefore likely to be within the statistical error, but also possibly can result in statistically significant subtle biases for large studies averaging over many molecules.  For high-precision calculations, these errors would be highly statistically significant, resulting in a lack of agreement and difficulties in reproducibility from one study to the next.

We also found that PME provides results that are consistent with potential switching but do not require an explicit switching function. While PME removes the dependence on cutoff-based artifacts, its additional computational cost and implementation complexity should be considered when performing free energy calculations.

%Although the effects of the $\lambda$ dependence of the energies at inside $r_c$ are not currently included in GROMACS, and likely other codes, it may be possible to include this lambda dependence in future calculations. It is not clear that the advantages of adding these additional corrections would be worth the code complexity added, given that that potential switching is cutoff independent already, but the exact tradeoffs are likely different for each software developer and user. 

What the ``right'' value of the cutoff distance or the choice of cutoff method is a point well worth discussion. Fundamentally, the choice of cutoff distance and the form that any switching function takes is a part of the force field itself, as it represents a change to the functional form. Thus, when choosing a particular force field to perform a simulation with, to be completely consistent, one should attempt to use the same cutoff distance and functional form that were used in the original parameterization. For example, when simulating lipid bilayers, the physical properties of the bilayer are significantly dependent on the cutoff.~\cite{Klauda_long_range,Huang_cutoffs_bilayer_2014,Wennberg:2013:J.Chem.TheoryComput.}. 

However, using the exact same cutoff distance and functional form as the original force field study is not always possible or desirable. Different choices in algorithms or computer architectures may alter the division of computing cost between long range and short range calculations, making it significantly faster to choose a different cutoff in some simulation engines and computing architectures than others. Sticking with the original cutoff would result in such cases in a heavy penalty in sampling efficiency on the new architecture and increase in statistical uncertainty that would more likely than offset any gains in accuracy by using the same cutoff.  In addition, different code bases may implement different functional forms for the shift or switch than each other, making it nearly impossible to directly replicate the energies generated when using different molecular dynamics software.~\cite{shirts_comparison_JCAMD_2016} It would be preferable therefore, if at all possible, to use methods whose physical properties are essentially independent of exactly how the division of short range and long range interactions are performed. Using PME, particle mesh, or other similar methods for dispersion interactions will work, regardless of the heterogeneity of the system.  For the purposes of computing free energies of solvation, and likely to some extent free energies of binding, then potential-switch works for high-precision calculations, whereas force-switch and potential-shift do not.

\section{Conclusions}
In this study, we investigated the influence of three van der Waals modifiers on the accuracy of free energy calculations. Our findings highlight the differences in how each modifier handles chosen cutoff induced discontinuities and their impact on the calculated free energy.
Analysis of potential energies as a function of distance with two Lennard-Jones spheres associated with each modifier confirmed correct implementation of the modifiers (and in one case led to key bug fixes). Subsequent analysis of calculated $\Delta G_{solv}$ showed the calculated values for force-switch and potential-shift modifiers are shifted positive, as attractive contributions to the free energy towards the solvent are being neglected, an effect that, as expected, becomes negligible at longer cutoff distances. 

This study conclusively demonstrates that the choice of modifier and cutoff can significantly impact the accuracy of free energy calculations, at least in current implementations in GROMACS and LAMMPS. The short-term action suggested by this study is that for GROMACS the potential-switch modifier should be used, and that the force-switch and potential shift modifiers should not be used in free energy simulations that require high precision. For LAMMPS, unshifted potentials including a tail correction are likely to produce the more accurate free energies,unless one uses a force field parameterized to unshifted potentials, though the lack of switching function introduces some small amount of error due to the lack of energy conservation. Additionally, while the use of PME for dispersion interactions also eliminates this cutoff dependence, with the added advantage that it is more transferrable to less homogeneous systems, its increased computational cost should be considered. Longer term, other approaches may be developed that give the correct independence of cutoff of free energy calculations. 

\section*{Author Contributions}
L.M.W., S. S., and M.R.S. conceptualized the project and designed the methodology. Experiments were performed and analyzed by L.M.W., with the exception of LAMMPS simulations, which were carried out by Y.R.. L.M.W. wrote the original manuscript draft; editing and review of the manuscript was done by M.R.S. and S.S.. M.R.S. and S.S. supervised the project and obtained resources.

\begin{acknowledgement}
We thank Berk Hess for discussions of current implementations of cutoffs in GROMACS and with rapid fixing of bugs found in the GROMACS force-switch free energy implementation. L. W. was supported in part by the NSF NRT Integrated Data Science Fellowship (award 2022138), and grant R01GM123296 from the National Institutes of General Medical Sciences. This work utilized the Alpine high performance computing resource at the University of Colorado Boulder. Alpine is jointly funded by the University of Colorado Boulder, the University of Colorado Anschutz, and Colorado State University.  S.S. acknowledges support of the National Science Foundation (NSF) CAREER grant 2046095. Computational resources for the LAMMPS simulations were provided by the NSF ACCESS grant DMR190005 and NSF MRI grant 2320493.  M.R.S has received an Open Science Fellowship from Psivant Therapeutics, and consults for Relay Therapeutics.
\end{acknowledgement}

\clearpage
\bibliography{citations/Software,citations/longrange}

\newpage

\newcommand{\ra}[1]{\renewcommand{\arraystretch}{#1}}
\renewcommand{\thetable}{S\arabic{table}}
\renewcommand{\thefigure}{S\arabic{figure}}
\renewcommand{\thesection}{S\arabic{section}}
\renewcommand{\thepage}{S\arabic{page}}

\begin{center}
{\Large Supporting information for ``Force switching and potential shifting lead to significant cutoff dependence in alchemical free energies''}\\
Lindsey M. Whitmore, Yalda Ramezani, Sumit Sharma, and Michael R. Shirts
\end{center}

\begin{figure}
    \centering
    \includegraphics[width=0.95\linewidth]{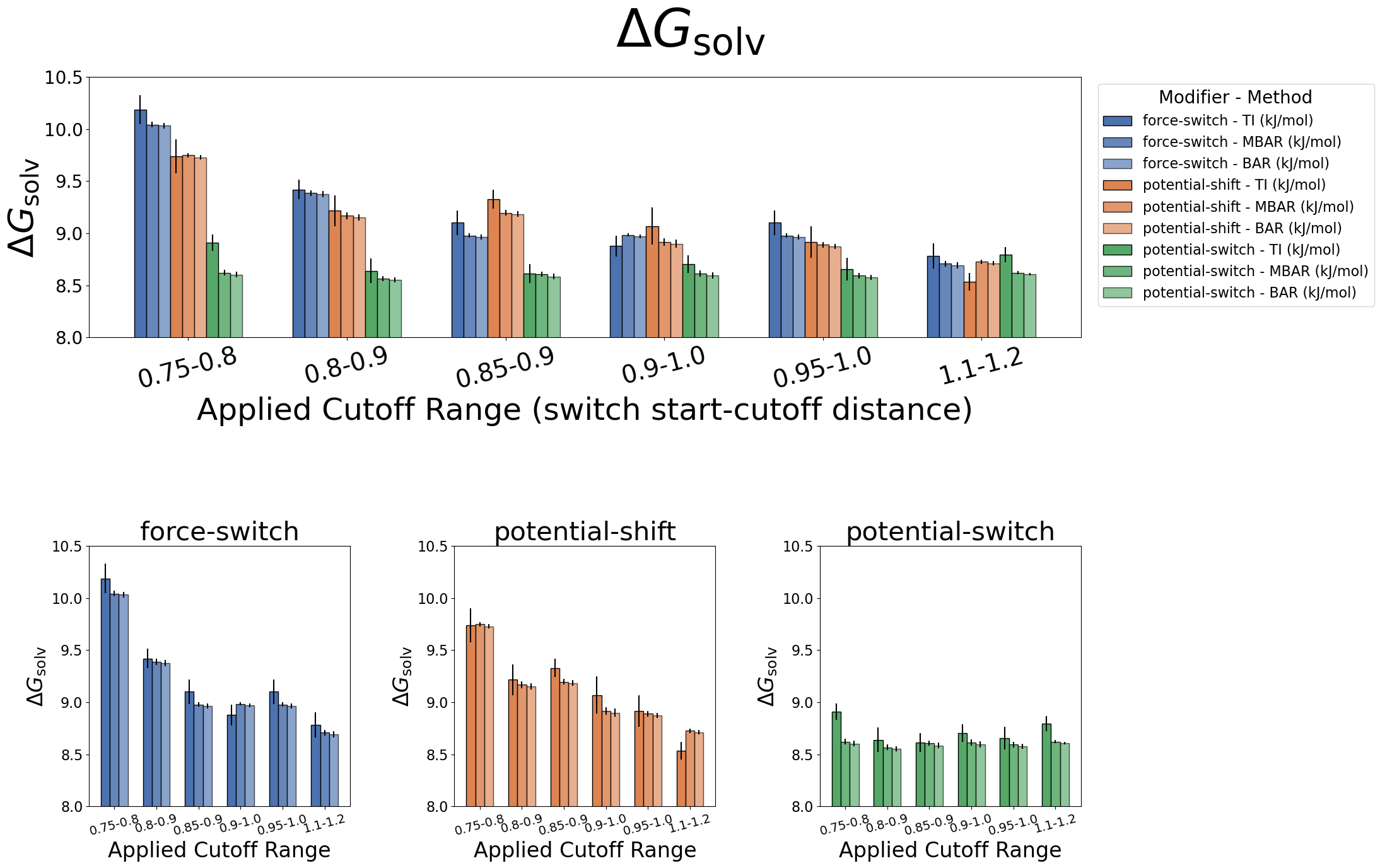}
    \caption{Free energy of solvation ($\Delta G_{solv}$) as a function of applied cutoff range. The top panel shows $\Delta G_{solv}$ across cutoff ranges for the force-switch (blue), potential-shift (orange), and potential-switch (green) modifiers, evaluated using the Multistate Bennett Acceptance Ratio (MBAR), Bennett Acceptance Ratio (BAR), and Thermodynamic Integration (TI) methods. $\Delta G_{solv}$ calculated with force-switch and potential-shift modifiers has significant cutoff dependence, whereas the $\Delta G_{solv}$ calculated with the potential-switch modifiers is essentially cutoff dependent. The bottom panels show the free energy contributions for each modifier individually. Error bars represent the standard error of the mean for $\Delta G_{solv}$ values over 10 replicates for each modifier across the applied cutoff ranges. MBAR and BAR are consistent to within statistical error of each other, while TI calculations are slightly higher, presumably because the spacing in $\lambda$ was insufficient to properly integrate $\langle dH/d\lambda\rangle$ with minimal bias given the size of the molecule and thus the relatively large curvature in $\langle dH/d\lambda \rangle$.}
    \label{SIfig:all_methods_FE}
\end{figure}

\newpage

\begin{table}[!ht]
    \centering
    \begin{tabular}{|c|c|c|c|c|c|c|c|}
    \hline
        Modifier & Cutoff & TI & TI SEM & MBAR & MBAR SEM & BAR & BAR SEM \\ \hline
        Potential-switch & 0.75-0.8 & 8.908 & 0.083 & 8.621 & 0.030 & 8.602 & 0.030 \\ \hline
        Potential-switch & 0.8-0.9 & 8.638 & 0.124 & 8.566 & 0.027 & 8.552 & 0.025 \\ \hline
        Potential-switch & 0.85-0.9 & 8.611 & 0.097 & 8.605 & 0.026 & 8.584 & 0.028 \\ \hline
        Potential-switch & 0.9-1.0 & 8.705 & 0.088 & 8.613 & 0.031 & 8.597 & 0.032 \\ \hline
        Potential-switch & 0.95-1.0 & 8.653 & 0.114 & 8.593 & 0.027 & 8.575 & 0.027 \\ \hline
        Potential-switch & 1.1-1.2 & 8.792 & 0.076 & 8.621 & 0.015 & 8.607 & 0.014 \\ \hline
        Potential-shift & 0.75-0.8 & 9.738 & 0.175 & 9.747 & 0.022 & 9.726 & 0.022 \\ \hline
        Potential-shift & 0.8-0.9 & 9.216 & 0.156 & 9.166 & 0.033 & 9.150 & 0.033 \\ \hline
        Potential-shift & 0.85-0.9 & 9.327 & 0.093 & 9.195 & 0.028 & 9.183 & 0.029 \\ \hline
        Potential-shift & 0.9-1.0 & 9.068 & 0.189 & 8.915 & 0.040 & 8.898 & 0.041 \\ \hline
        Potential-shift & 0.95-1.0 & 8.916 & 0.158 & 8.889 & 0.027 & 8.873 & 0.028 \\ \hline
        Potential-shift & 1.1-1.2 & 8.534 & 0.091 & 8.726 & 0.023 & 8.712 & 0.024 \\ \hline
        Force-switch & 0.75-0.8 & 10.187 & 0.148 & 10.043 & 0.030 & 10.033 & 0.029 \\ \hline
        Force-switch & 0.8-0.9 & 9.420 & 0.100 & 9.384 & 0.031 & 9.372 & 0.032 \\ \hline
        Force-switch & 0.85-0.9 & 9.101 & 0.126 & 8.978 & 0.024 & 8.963 & 0.026 \\ \hline
        Force-switch & 0.9-1.0 & 8.876 & 0.103 & 8.984 & 0.018 & 8.970 & 0.017 \\ \hline
        Force-switch & 0.95-1.0 & 9.101 & 0.126 & 8.978 & 0.024 & 8.963 & 0.026 \\ \hline
        Force-switch & 1.1-1.2 & 8.781 & 0.129 & 8.708 & 0.030 & 8.692 & 0.030 \\ \hline
        PME no modifier & 0.9 & 8.772 & 0.099 & 8.640 & 0.022 & 8.626 & 0.022 \\ \hline
        PME potential-shift & 0.9 & 8.535 & 0.091 & 8.633 & 0.021 & 8.622 & 0.021 \\ \hline
    \end{tabular}
    \caption{Tabular version of the data presented in Figure \ref{SIfig:all_methods_FE} with additional PME data in units of kJ/mol. Uncertainties shown as error bars in Figure \ref{SIfig:all_methods_FE} calculated from standard error in the mean from 10 replicates.\label{SItable:all_methods_FE
    }
    }
\end{table}

\begin{table}[h]
    \centering
    \begin{tabular}{|ccc|}
    \hline
        state & $\lambda$ & weight ($k_BT$)\\
        \hline
0 &0.0 & 0 \\
1& 0.09 & 2.19552\\
2&0.18 & 4.32298 \\
3& 0.30  & 6.95010\\
4&0.42  & 9.25799\\
5&0.495  & 10.48891\\
6&0.57 & 11.45557 \\
7&0.625 & 11.89813\\
8&0.68 & 11.88643\\
9&0.72 & 11.43653\\
10&0.76 & 10.43653 \\
11&0.81 & 8.38093\\
12&0.86 & 6.22252\\
13&0.93 & 4.35859\\
14 &1.0 & 3.39715 \\
        \hline
\end{tabular}
\caption{Fixed initial weights for expanded ensemble simulations of anthracene in water}
\label{SI:tab:ee_weights}
\end{table}

\begin{table}[h]
    \centering
    \begin{tabular}{lcc}
    \hline
        \textbf{Alkane} & \textbf{Initial Box Length (\AA)} & \textbf{Number of Water Molecules} \\
        \hline
        $C_1$--$C_8$  & 32  & 1500 \\
        $C_9$--$C_{14}$ & 42  & 3390 \\
        $C_{15}$--$C_{20}$ & 52  & 6440 \\
        \hline
    \end{tabular}
    \caption{Nominal simulation box length and number of water molecules in the simulation systems for LAMMPS for different alkanes.}
    \label{SI:tab:alkane_water}
\end{table}

\begin{table}[!ht]
    \centering
    \begin{tabular}{|c|c|c|c|c|c|c|c|}
    \hline
        Modifier & Cutoff & TI & TI Error & MBAR & MBAR Error & BAR & BAR Error \\ \hline
        Potential-switch & 0.75-0.8 & 8.908 & 0.350 & 8.621 & 0.037 & 8.602 & 0.028 \\ \hline
        Potential-switch & 0.8-0.9 & 8.638 & 0.351 & 8.566 & 0.034 & 8.552 & 0.026 \\ \hline
        Potential-switch & 0.85-0.9 & 8.611 & 0.346 & 8.605 & 0.036 & 8.584 & 0.027 \\ \hline
        Potential-switch & 0.9-1.0 & 8.705 & 0.348 & 8.613 & 0.034 & 8.597 & 0.026 \\ \hline
        Potential-switch & 0.95-1.0 & 8.653 & 0.351 & 8.593 & 0.034 & 8.575 & 0.026 \\ \hline
        Potential-switch & 1.1-1.2 & 8.792 & 0.345 & 8.621 & 0.034 & 8.607 & 0.026 \\ \hline
        Potential-shift & 0.75-0.8 & 9.738 & 0.371 & 9.747 & 0.036 & 9.726 & 0.027 \\ \hline
        Potential-shift & 0.8-0.9 & 9.216 & 0.359 & 9.166 & 0.036 & 9.150 & 0.027 \\ \hline
        Potential-shift & 0.85-0.9 & 9.327 & 0.353 & 9.195 & 0.034 & 9.183 & 0.026 \\ \hline
        Potential-shift & 0.9-1.0 & 9.068 & 0.439 & 8.915 & 0.042 & 8.898 & 0.032 \\ \hline
        Potential-shift & 0.95-1.0 & 8.916 & 0.355 & 8.889 & 0.036 & 8.873 & 0.027 \\ \hline
        Potential-shift & 1.1-1.2 & 8.534 & 0.345 & 8.726 & 0.034 & 8.712 & 0.026 \\ \hline
        Force-switch & 0.75-0.8 & 10.187 & 0.375 & 10.043 & 0.035 & 10.033 & 0.026 \\ \hline
        Force-switch & 0.8-0.9 & 9.420 & 0.365 & 9.384 & 0.035 & 9.372 & 0.026 \\ \hline
        Force-switch & 0.85-0.9 & 9.101 & 0.348 & 8.978 & 0.036 & 8.963 & 0.027 \\ \hline
        Force-switch & 0.9-1.0 & 8.876 & 0.352 & 8.984 & 0.034 & 8.970 & 0.026 \\ \hline
        Force-switch & 0.95-1.0 & 9.101 & 0.348 & 8.978 & 0.036 & 8.963 & 0.027 \\ \hline
        Force-switch & 1.1-1.2 & 8.781 & 0.419 & 8.708 & 0.042 & 8.692 & 0.031 \\ \hline
        PME no modifier & 0.9 & 8.772 & 0.349 & 8.640 & 0.036 & 8.626 & 0.027 \\ \hline
        PME potential-shift & 0.9 & 8.535 & 0.353 & 8.633 & 0.034 & 8.622 & 0.026 \\ \hline
    \end{tabular}
    \label{table:free_energy_averages}
    \caption{Average $\Delta G$ and error for TI, BAR, and MBAR in units of kJ/mol. Here, BAR error was propagated using $\sqrt{\sum_i (\Delta G_i)^2}$, where individual $\Delta G_i$ values correspond to uncertainties from each lambda interval. While this approach provides a reasonable estimate, it is known to slightly underestimate the actual uncertainty in some cases.}
\end{table}

\begin{table}[h!]
\begin{center}
\caption{Hydration free energy of alkanes calculated using the free energy perturbation (FEP) methodology using shifted and unshifted potentials. The results are compared with experimental values and group contribution method. All hydration free energy values are in kJ/mol. FEP$_{01}$ and FEP$_{10}$ refer to the forward and reverse paths, respectively.}
\begin{tabular}{c|cc|cc|c|c}
\toprule
\textbf{Alkane} & \multicolumn{2}{c|}{\textbf{Unshifted}} & \multicolumn{2}{c|}{\textbf{Shifted}} & \textbf{Experimental} & \textbf{Group Cont.} \\
 & FEP$_{01}$ & FEP$_{10}$ & FEP$_{01}$ & FEP$_{10}$ & & \\
\midrule
1  & 8.62  & 8.63  & 8.50  & 8.45  & 8.37       &         \\
2  & 7.89  & 8.00  & 7.87  & 8.11  & 7.66       &         \\
3  & 8.55  & 8.65  & 8.91  & 8.85  & 8.18       &         \\
4  & 8.99  & 9.06  & 9.59  & 9.49  & 8.70       &         \\
5  & 9.90  & 10.13 & 9.97  & 10.22 & 9.76       &         \\
6  & 11.07 & 10.59 & 11.18 & 11.06 & 10.40      &         \\
7  & 11.37 & 11.15 & 11.77 & 12.00 & 10.96      &         \\
8  & 11.97 & 11.70 & 12.80 & 13.19 & 12.10      &         \\
9  & 12.69 & 12.97 & 13.88 & 13.97 &            & 12.58   \\
10 & 13.38 & 13.81 & 14.72 & 14.80 &            & 13.32   \\
11 & 14.62 & 14.31 & 15.63 & 15.77 &            & 14.06   \\
12 & 15.50 & 15.29 & 16.99 & 17.22 &            & 14.80   \\
13 & 16.39 & 16.10 & 16.88 & 17.34 &            & 15.54   \\
14 & 16.92 & 16.43 & 17.94 & 18.61 &            & 16.28   \\
15 & 16.87 & 17.94 & 19.11 & 19.76 &            & 17.02   \\
16 & 18.07 & 18.27 & 20.92 & 20.58 &            & 17.76   \\
17 & 18.37 & 19.33 & 21.29 & 21.09 &            & 18.50   \\
18 & 19.95 & 20.12 & 22.49 & 21.59 &            & 19.24   \\
19 & 21.55 & 21.55 & 23.07 & 23.13 &            & 19.98   \\
20 & 21.73 & 21.65 & 24.03 & 23.56 &            & 20.72   \\
\bottomrule \label{tab:SIlammps_free_energies}
\end{tabular}
\end{center}
\end{table}

\begin{table}[h]
    \centering
    \begin{tabular}{|c|c|}
    \hline
      Cutoff method   &  Speed (ns/day)\\
      \hline
      Potential shift (0.8--0.9 nm)  & 163 $\pm$ 6 \\ 
      Potential switch (0.8--0.9 nm) & 155 $\pm$ 10 \\
      Force switch (0.8--0.9 nm)     & 148 $\pm$ 2 \\
      PME vdw (0.9 nm, no modifier) &  98 $\pm$ 7  \\
      PME vdw (0.9 nm, potential shift) & 119 $\pm$ 7 \\
      \hline
    \end{tabular}
    \caption{Speed of the expanded ensemble free energy calculations in ns/day with various cutoff methods for calculations of the free energy of anthracene.  Uncertainties calculated as standard error of the mean over 10 replicates.}
    \label{tab:SIpme_speed}
\end{table}

\clearpage
%\begin{figure}
%    \centering
%    %\includegraphics[width=0.7\linewidth]{Figures/MTREXEE_TOC.png}
%    \caption{For Table of Contents Only}
%    \label{fig:enter-label}
%\end{figure}
\end{document}